# Transverse detuned loading for modulation bandwidth enhancement in VCSELs

Giacomo Graziano [1,2], Richard Schatz[3], Antoine Pissis[1], Wilfried Maineult[1], Aurelio Rossinelli[1], Philipp Täschler[1], Simon Hönl[1], Vincenzo Pasanisi[1], Tobias Süss[1], Dario Schiavon[1], Donato Bonfrate[1], Martin Spieser[1], Stefano Tirelli[1], Evgeny Zibik[1]

1. Coherent, II-VI Laser Enterprise GmbH, Binzstrasse 17, 8045 Zurich, Switzerland

2. James Watt School of Engineering, University of Glasgow, Glasgow, UK

3. Independent Researcher and Consultant, Danderyd, Sweden
*giacomo.graziano@coherent.com*

***Abstract*— Directly modulated vertical-cavity surface-emitting lasers (VCSELs) are central to short-reach optical interconnects, where modulation bandwidth and power efficiency are critical factors for increasing data transfer capacity. Conventional oxide VCSELs still require power-hungry signal processing to achieve high-speed data transmission above 100 Gb/s. Here we report an experimental implementation of detuned loading in a VCSEL to enhance the modulation bandwidth. The detuned loading is achieved by carrier-induced index variations that reshape an engineered transverse mode so its overlap with spatially selective optical losses decreases with increased carrier density. This produces a negative differential loss reminiscent of that in detuned-loaded DBR lasers lasing on the long-wavelength reflectivity flank. The resulting 850 nm single-mode VCSEL exhibits an enhanced carrier-photon relaxation frequency of 44 GHz, yielding a 3dB electrical modulation bandwidth exceeding 50 GHz. The enhancement persists from 25 °C to 75 °C showing the robustness of the mechanism introduced. The same device supports 106.25 Gb/s NRZ transmission without digital signal processing, transmitter pre-emphasis or receiver equalization. These results establish transverse detuned loading as a robust alternative approach for extending the modulation bandwidth of vertical-cavity surface-emitting lasers.**

## INTRODUCTION

The rapid growth of artificial-intelligence and high-performance-computing workloads is increasing the aggregate bandwidth required from short-reach interconnects. As optical interfaces move closer to the compute, bandwidth density, energy efficiency, footprint and achievable speed per lane become critical figures of merit [1] [2]. Several technologies address these requirements with different trade-offs [3] [4].

Silicon microring modulators provide very small footprints and low switching energies, but are sensitive to wavelength and temperature variations [5]. Mach–Zehnder modulators offer broadband, more temperature-robust operation, at the cost of larger footprints and higher drive requirements [6]. Electro-absorption modulators combine compactness and high speed but introduce insertion loss and wavelength sensitivity [7]. Pockels-based modulators offer high bandwidth, low optical loss and low drive voltage, although they generally require relatively long interaction lengths [8]. At shorter reaches, copper interconnects are increasingly constrained by reach and bandwidth density [9]. Overall, most of the transmitter power is still consumed by digital signal processing (DSP), equalization and thermal control, while receiver power increases with photodiode and transimpedance-amplifier bandwidth [10]. External or integrated modulators also introduce insertion loss that must be compensated by higher laser power.

VCSEL arrays are attractive for scale-up interconnects because they combine compactness, dense parallelization, direct modulation, high optical efficiency and broad-temperature operation without active stabilization [11] [12] [13]. With simple NRZ signalling, they can reduce DSP and equalization requirements while simplifying the link architecture and lowering power consumption. Conventional oxide-confined VCSELs have recently reached modulation bandwidths approaching 40 GHz [14] [15]. Operation at 106.25 Gb/s NRZ, however, requires higher intrinsic bandwidth, approaching the 53 GHz Nyquist frequency of 106.25 GBd signalling.

Bandwidth enhancement beyond the conventional carrier–photon resonance limit can address this gap by modifying the laser dynamics. One route is to equalize the response at frequencies beyond the photon-carrier relaxation frequency by using an additional optical cavity resonance close to the main mode. This photon-photon resonance can be achieved in VCSELs by using coupled cavities, or transverse-resonance effects [16] [17]. Using this scheme, optical bandwidth >50GHz has been reported in 1060 nm metal aperture VCSEL when enhanced by the dispersion of single mode fibre [18] [19]. However, this approach requires precise control of modal spacing and bias point, making performance sensitive to epitaxial tuning, processing tolerances, self-heating and temperature-induced wavelength shifts [20].

External or integrated modulators reduce dependence on the intrinsic carrier–photon resonance but add epitaxial, fabrication and integration complexity [21] [22]. High-contrast-grating VCSELs enable short effective cavities and advanced mirror engineering for tunability and polarization control [23], but high-speed datacom implementation remains challenging because of fabrication tolerances, optical loss and thermal management [24] [25]. In edge-emitting lasers, detuned loading enhances bandwidth by positioning the lasing mode on a wavelength-dependent cavity-loss slope. Carrier-induced frequency chirp is thereby converted into dynamic loss modulation, modifying the damping and reinforcing the intrinsic response [26] [27] [28]. Unlike photon-photon resonance, detuned loading does not require higher-order mode coupling and may therefore offer improved stability with current and temperature. For VCSELs it was suggested that an external cavity could induce detuned loading [29] [30]. Here, we experimentally realize transverse detuned loading through the interaction between an engineered cavity mode and spatially selective loss, without relying on resonant coupling to another cavity. We demonstrate an 850-nm single-mode VCSEL with a directly resolved carrier–photon resonance near 44 GHz and an electrical bandwidth beyond the 50 GHz measurement range. The enhancement persists from 25 °C to 75 °C, while the device supports 106.25 Gb/s NRZ modulation without DSP, transmitter pre-emphasis or receiver equalization. These results establish transverse detuned loading as a viable route for bandwidth enhancement in vertical-cavity lasers.

## DETUNED-LOADING EFFECT

The concept of detuned loading in lasers "involves the introduction of any dispersive loss mechanism into the laser followed by deliberate detuning of the lasing frequency away from the loss peak" [26]. It has mainly been realized and analysed in edge-emitting lasers, especially DBR lasers [31] [28] [32] [33] but the underlying mechanism is not limited to a one-dimensional geometry. It can be extended to an arbitrary three-dimensional open optical cavity using the general theory of non-Hermitian quasi-mode dynamics by Wenzel and Wünsche [34]. In this formulation, the optical field is expanded into the instantaneous complex eigenmodes of the cavity. Carrier-induced changes in the complex propagation constant dynamically perturb both the complex eigenfrequencies and the spatial modal intensity distributions. Following [34] and [35], the total photon number S and the complex reactive photon number U of a single-mode cavity can be expressed in the slowly varying electric and magnetic peak fields $\mathbf{E_0}(\mathbf{r},t)$ and $\mathbf{H_0}(\mathbf{r},t)$ of the lasing mode as:

$$S = \frac{1}{4\hbar\omega}\int_V \left[\varepsilon_0 Re\left(\frac{\partial\,\tilde{\omega}\varepsilon}{\partial\tilde{\omega}}\right)|\mathbf{E_0}|^2 + \mu_0|\mathbf{H_0}|^2\right] dV \qquad (1)$$

$$U = \frac{1}{4\hbar\omega}\int_V \left[\varepsilon_0 \frac{\partial\,\tilde{\omega}\varepsilon}{\partial\tilde{\omega}}\mathbf{E_0}\cdot\mathbf{E_0} - \mu_0\mathbf{H_0}\cdot\mathbf{H_0}\right] dV \qquad (2)$$

The ratio between S and |U| is given by the Petermann factor $K = (S/|U|)^2$. In a closed cavity without net loss, the electric and magnetic fields are 90 degrees out of phase, resulting in the two integrals being equal ($S = U$) and a Petermann factor of exactly 1. Conversely, in a reflection-free waveguide without net loss, the electric and magnetic fields of the traveling wave are in phase, causing the integral $U$ to evaluate to zero. One can also define a total confinement factor, Γ, and a reactive confinement factor, $\tilde{\Gamma}$, as the relative fractions of $S$ and $U$ in the active layer, respectively. The reactive confinement can alternatively be expressed as the sensitivity of the complex modal eigenfrequency, $\tilde{\omega}$, to a gain and/or index perturbation in the active layer [32]:

$$\tilde{\Gamma} \equiv \frac{1}{4\hbar\omega U}\int_a \left[\varepsilon_0 \frac{\partial\,\tilde{\omega}\varepsilon}{\partial\tilde{\omega}}\mathbf{E_0}\cdot\mathbf{E_0} - \mu_0\mathbf{H_0}\cdot\mathbf{H_0}\right] dV =$$

$$= -\frac{\lambda}{2\pi v_g}\frac{\Delta\tilde{\omega}}{\Delta\tilde{n}_a} \qquad (3)$$

where $\Delta\tilde{n}_a = \Delta n_a + \mathrm{i}\Delta\kappa_a$ is the complex refractive index perturbation in the active layer. The reactive confinement factor will hence have an imaginary part if a change in the real part of the refractive index induces a change in $\mathrm{Im}(\tilde{\omega}) = (\gamma - G)/2$, i.e., the net modal loss rate. Here, $\gamma$ is the cavity loss and $G = \Gamma v_g g$ is the cavity gain where $g$ is the material gain and $v_g$ is the group velocity. The carrier-photon relaxation frequency of the laser, neglecting gain compression, can be written [32]:

$$f_r = \frac{1}{2\pi}\sqrt{\frac{S}{\tau_p V_a} v_g \frac{dg}{dN}\mathrm{Re}\left[(1+i\alpha)\tilde{\Gamma}\right]} \qquad (4)$$

where $S$ is the photon number, $N$ is the carrier density, $V_a$ is the volume of active layer and $\tau_p$ is the photon lifetime. The relaxation frequency is mainly determined by the photon density and the differential gain, $dg/dN$ of the active material. However, it is seen that the relaxation frequency is also affected by the linewidth enhancement factor, $\alpha$, if the reactive confinement factor has an imaginary part. Since α is positive, the relaxation frequency will be enhanced if the cavity is engineered such that $\mathrm{Im}(\tilde{\Gamma}) < 0$. This happens when the net modal losses are increased with an increased real part of the refractive index, corresponding to a reduced carrier density in the active region, giving:

$$\frac{\partial \mathrm{Im}(\tilde{\omega})}{\partial n_a} > 0 \qquad (5)$$

An increase in the carrier density will then generate negative differential loss that increases the effective differential gain and relaxation frequency. Similarly, the damping without gain compression can be derived as [32]:

$$\gamma_d = \frac{1}{\tau_d} + \frac{S}{\tau_p V_a}\frac{d}{dN}\ln\left(g\Gamma\sqrt{K}\right) \qquad (6)$$

where $\tau_d$ is the differential carrier lifetime. Although the Petermann factor $K$ is close to 1 in a laser with low outcoupling losses like a VCSEL, the logarithmic derivative $d\ln(\Gamma\sqrt{K})/dN$ can still be significant and negative. When this happens, the damping is reduced, which can even lead to self-pulsations [33] [36] [32]. The combination of increased relaxation frequency

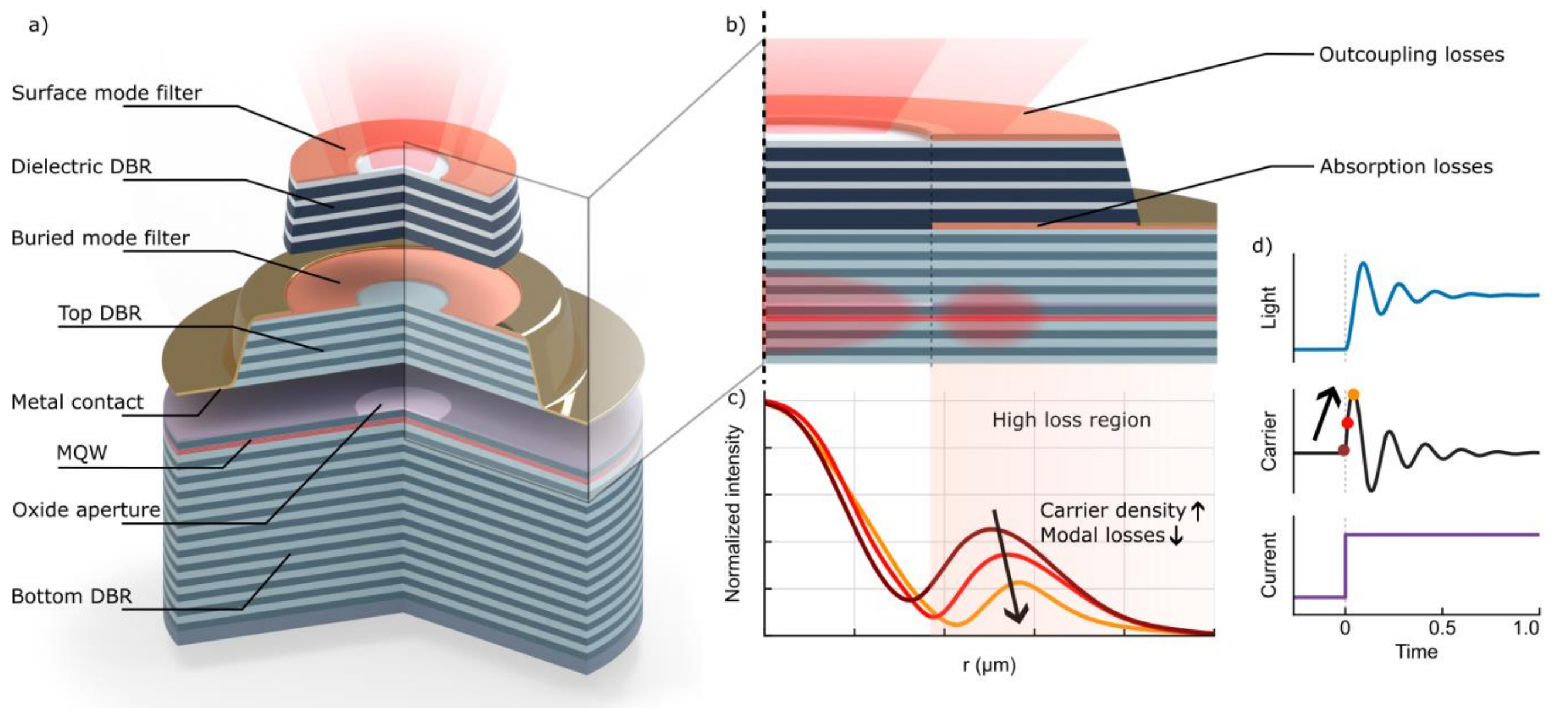

*Figure 1 a) Schematics of the exploded structure of a detuned-loading VCSEL with description of the main elements. b) Vertical cross-section of the structure. The engineered mode is overlaid, and the two sources of losses are indicated. (c) Cross-sections of the simulated optical mode within the active-region plane of a detuned-loading VCSEL for different MQW refractive indices, representing different carrier densities. Higher carrier density is depicted in orange and lower in dark red. d) Schematic picture of light intensity, carrier density and current in the MQW as a function of time. The carrier densities corresponding to the curves simulated in c) are highlighted with the same colours.*

and undamping can therefore provide a strong enhancement of the modulation bandwidth of the device.

## Detuned-Loading VCSEL Design

Bandwidth enhancement through detuned loading has been experimentally demonstrated for DBR and DR semiconductor lasers, where the optical mode is lasing on the long wavelength flank of the dispersive Bragg reflector [28] [33] [37]. Implementing the detuned-loading effect in a VCSEL by positioning the lasing resonance on a steep flank of the DBR reflectivity spectrum is impractical, as it would require precise alignment of the cavity resonance, gain spectrum, and mirror response [38]. Moreover, this approach would make the effect highly sensitive to bias and temperature variations. We therefore implement detuned loading through the transverse degree of freedom. Carrier-density fluctuations modify the refractive index of the MQW region and consequently perturb the transverse field distribution. By engineering the cavity so that this field redistribution changes the overlap with spatially selective loss, refractive-index modulation is converted into modal-loss modulation without relying on spectral detuning. In conventional VCSELs, the optical-loss distribution is generally nearly uniform across the transverse plane, so carrier- and temperature-induced changes in transverse confinement are not expected to produce a strong variation in modal loss.

To modify this behaviour, we introduced an intracavity surface relief inside the contact layer as depicted in Figure 1. Details of the full structure are reported in the Methods section. Modifying the cavity's transverse effective refractive index profile excites an engineered transverse mode whose field extends beyond the central region of the aperture. This mode exhibits a transverse node, with its intensity distribution split between a central lobe and an outer ring. This engineered mode is combined with additional spatially selective losses that partially overlap with the outer ring of the optical field: absorption losses from the contact layer and mirror losses from the silicon antiphase cap.

Figure 1 shows the simulated mode intensity profile superimposed on the schematic of the VCSEL structure. Most of the optical intensity remains localized inside the oxide aperture, where the cavity stack provides the lowest optical losses. However, unlike a standard Gaussian-like fundamental mode, a non-negligible fraction of the engineered mode overlaps with a high-loss region defined by the absorbing contact layer and the surface relief.

To quantify the detuned loading effect, we calculate the transverse eigenmodes while varying the refractive index of the active MQW region; simulation details are given in Methods. The sweep captures the bias-dependent evolution of the steady-state mode, while the local modal-loss slope around each operating point determines the response to the much smaller carrier-induced index modulation.

The simulation results show that the engineered mode changes shape as the MQW refractive index varies. In particular, the optical intensity in the outer region of the cavity changes with the refractive-index perturbation. Because this outer region partially overlaps with spatially selective losses, the mode-profile variation is converted into a variation of modal losses.

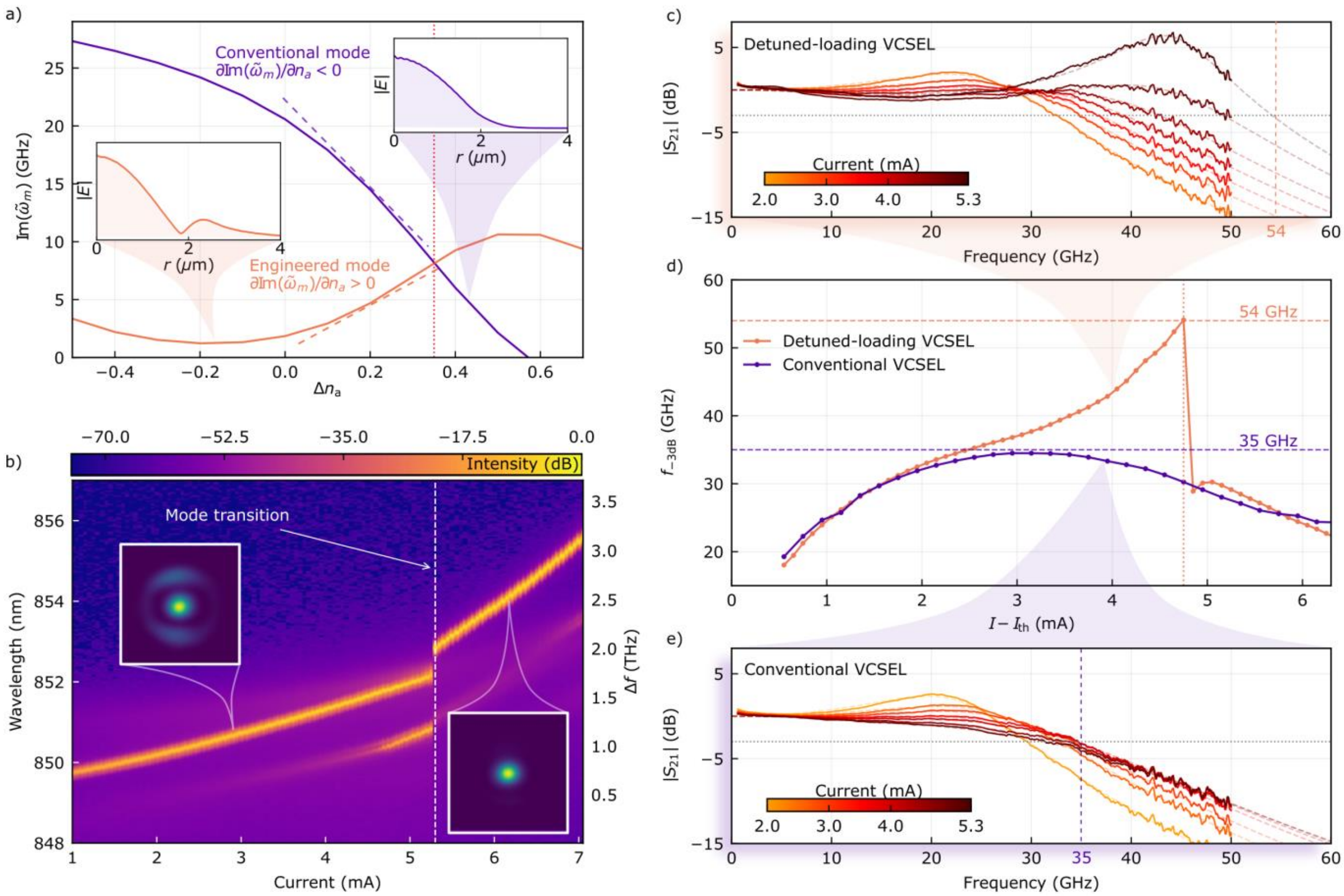

*Figure 2 a) Imaginary part of the modal frequency (net cavity loss) of the engineered mode and of the fundamental mode with respect to the variation of the pumped active area refractive index, showing positive and negative first order derivative. The insets show mode profiles extracted at the MQW before and after the mode transition. b) Spectral evolution of a detuned-loading VCSEL with respect to current. Measured near-field images of a detuned-loading VCSEL before and after the mode transition are displayed in the insets. c) The measured modulation response of the detuned-loading VCSEL and e) conventional VCSEL at different bias currents. Solid lines are experimental data, and dashed lines are the corresponding fits to a three poles transfer function. d) 25°C small signal modulation bandwidth comparison between a conventional VCSEL (purple) and a detuned-loading VCSEL (orange) with respect to current above threshold. The thresholds of the conventional VCSEL and the detuned-loading VCSEL are 0.45 mA and 0.55 mA, respectively.*

## EXPERIMENTAL RESULTS

### *Modal Evolution versus Bias*

Figure 2 compares the simulation results with the experimental characterization of a fabricated detuned-loading VCSEL. It depicts the modal losses as a function of the MQW refractive-index variation for the two modes with the lowest losses. One is the engineered mode that extends in the lossy region, and the other is a conventional Gaussian mode.

As seen from Eq. 5, the bandwidth enhancement is determined by the slope of the modal-loss curve shown in Figure 2a. Only the engineered mode exhibits a positive slope of modal loss versus MQW refractive-index variation. In contrast, the conventional mode shows a negative slope. The opposite sign of the response of the engineered mode is therefore the key feature that enables detuned loading in the VCSEL.

The measured spectral evolution of the device is shown in Figure 2b. The detuned-loading VCSEL is predominantly single mode. The near-field image of the lasing mode confirms that the experimentally observed mode corresponds to the simulated engineered mode. However, above 5.3 mA, a mode hop occurs and the laser switches to the standard Gaussian-like fundamental mode. A similar behaviour is observed in the simulations: as the refractive-index profile changes, the cavity increasingly favours the conventional mode, which suddenly becomes the one with the lowest losses. Also in this case, the simulated mode in Figure 2a corresponds to the measured one in Figure 2b.

### *Small-signal Modulation*

Figure 2d compares the measured small-signal modulation response of a detuned-loading VCSEL with the closest conventional single-mode reference VCSEL from the same wafer. Co-fabrication minimizes variations in growth, thickness, and processing, isolating the cavity design as the principal distinction: the reference uses a surface relief, whereas the detuned-loading device incorporates an intracavity mode filter and spatial loss design. Both aperture diameters are estimated to be around 3 μm. At 25 °C, the reference bandwidth saturates near 35 GHz, while the detuned-loading response

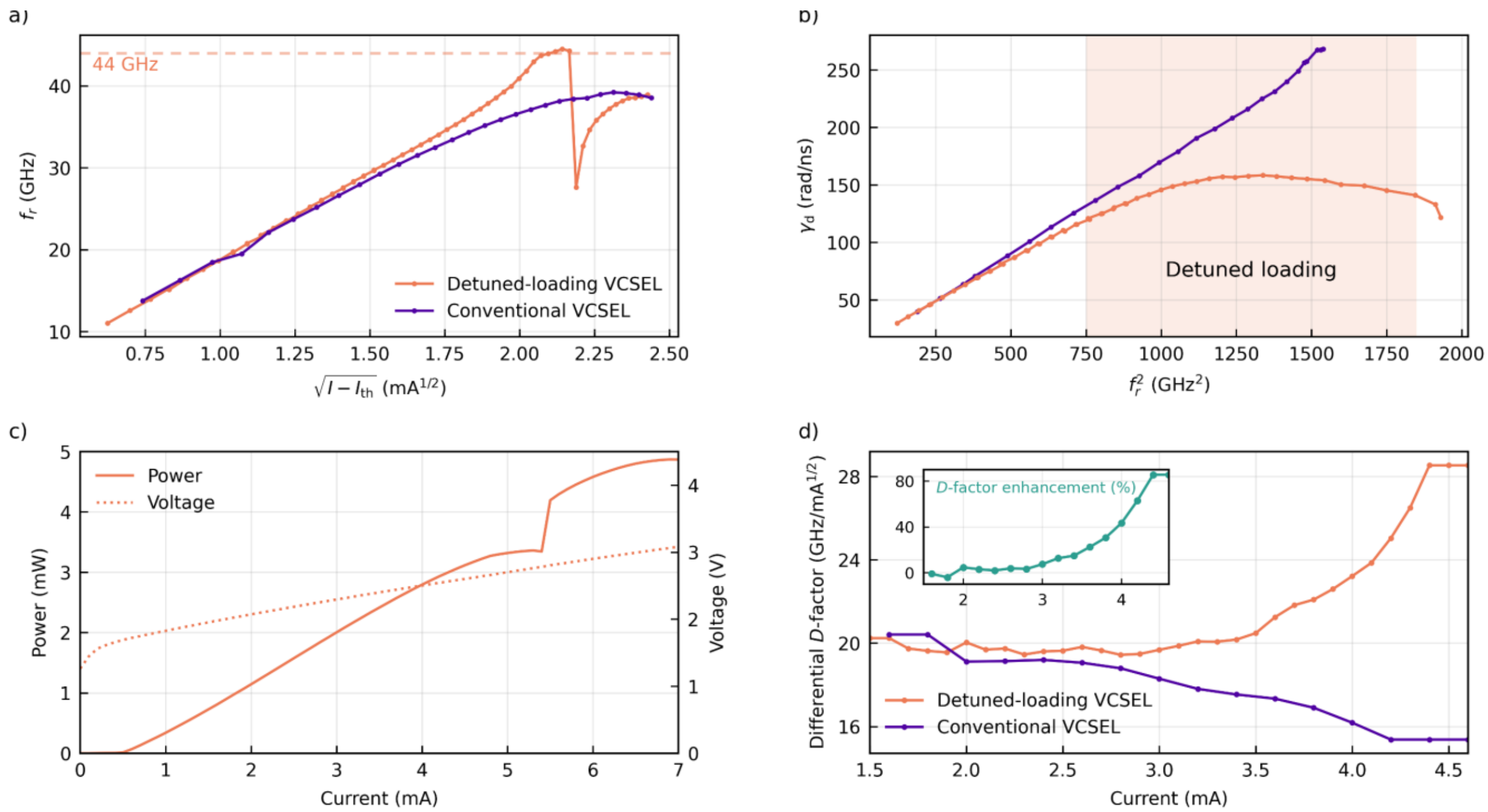

*Figure 3 a) Resonance frequency behaviour for a detuned-loading VCSEL and a conventional VCSEL with respect to $\sqrt{I - Ith}$ at 25°C. b) Relaxation damping versus the square of the relaxation frequency. c) Light-current and voltage-current characteristics of the detuned- loading VCSEL. The highlighted region corresponds to the range where the detuned-loading VCSEL shows a distinct difference with respect to the conventional device performance. d) Extracted differential D-factor from a) as a function of bias current. The inset displays the increase in the differential D-factor in the detuned-loading VCSEL. The increase is defined as the ratio between the differential D-factor in the two devices.*

remains above the -3 dB level at the 50 GHz measurement limit (curve fitting is used to estimate the crossing point beyond this range; see Methods). This enhancement persists across the entire bias range preceding the transverse-mode transition. At ~3.5 mA above threshold, where the reference approaches peak bandwidth, the detuned-loading device exceeds 40 GHz and continues rising until the mode transition. The small signal modulation exhibits a single relaxation peak shifting smoothly with current, without additional peaks or high-frequency shoulders. The device remains single-mode, with a frequency spacing exceeding 500 GHz between the lasing mode and the nearest higher-order mode. This confirms that the bandwidth enhancement reflects a modified fundamental carrier-photon response rather than a photon-photon resonance. Figure 3a explicitly highlights this distinction: both devices scale similarly at low currents, but diverge around 3 mA above threshold. The detuned-loading VCSEL relaxation frequency climbs past 44 GHz, whereas the conventional VCSEL saturates earlier. Figure 3b shows the accompanying increase in the relaxation damping. While the damping of the conventional VCSEL increases, as expected, linearly with the square of the relaxation frequency $f_{res}^2$, the detuned-loading VCSEL exhibits a sublinear behaviour that reverses to a decrease of damping before the mode transition; reminiscent of the behaviour in detuned-loading DBR lasers [33]. The simultaneous increase in $f_{res}$ and reduction in damping yields an electrical-electrical modulation bandwidth exceeding 50 GHz at 5.3 mA with a clearly visible 44 GHz relaxation peak in the measured S21 response. Figure 3c shows the LIV curve of the detuned-loading device. The threshold current is 0.55 mA and the output power reaches above 3 mW in the bias region affected by the detuned loading. Figure 3d tracks modulation efficiency versus applied current. The extracted differential D-factor in the detuned-loading VCSEL rises with current above 28 GHz/$\sqrt{\text{mA}}$, whereas the reference differential D-factor steadily decreases due to thermal effects. The inset shows the relative D-factor enhancement, defined as the ratio between the differential D-factors of the detuned-loading and conventional VCSELs. The current-induced change in the refractive index alters the overlap of the optical field with the patterned high-loss regions, resulting in a bias-dependent differential modal loss. Experimentally, this produces an enhancement of the differential D-factor that approaches a factor of two. Below the mode transition, the laser operates in the engineered mode with a favourable modal-loss derivative; above it, the device switches to a standard mode, collapsing the D-factor and bandwidth to that of the conventional VCSEL. Figure 4 demonstrates thermal robustness. From 25 °C to 75 °C, the reference bandwidth drops from 35 GHz to 30 GHz, while the detuned-loading device maintains its low-damping dynamics, reaching >50 GHz (25 °C), 48 GHz (50 °C), and 41 GHz (75 °C).

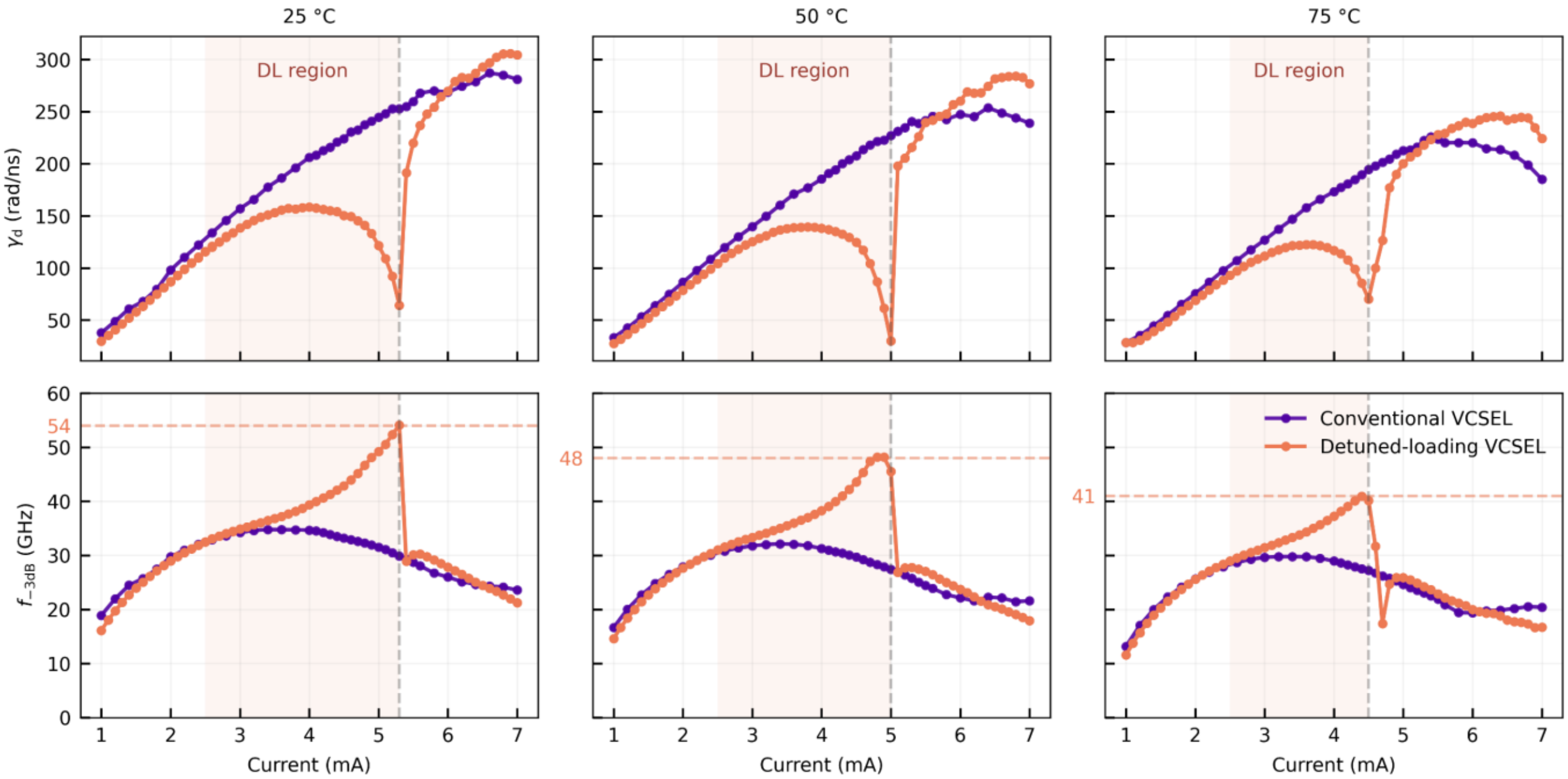

*Figure 4 a-c) Damping and d-f) -3dB bandwidth as a function of current for three heatsink temperatures: 25°C, 50°C and 75° for a detuned-loading VCSEL and a conventional VCSEL. Light orange overlay highlights the region interested by the detuned-loading effect.*

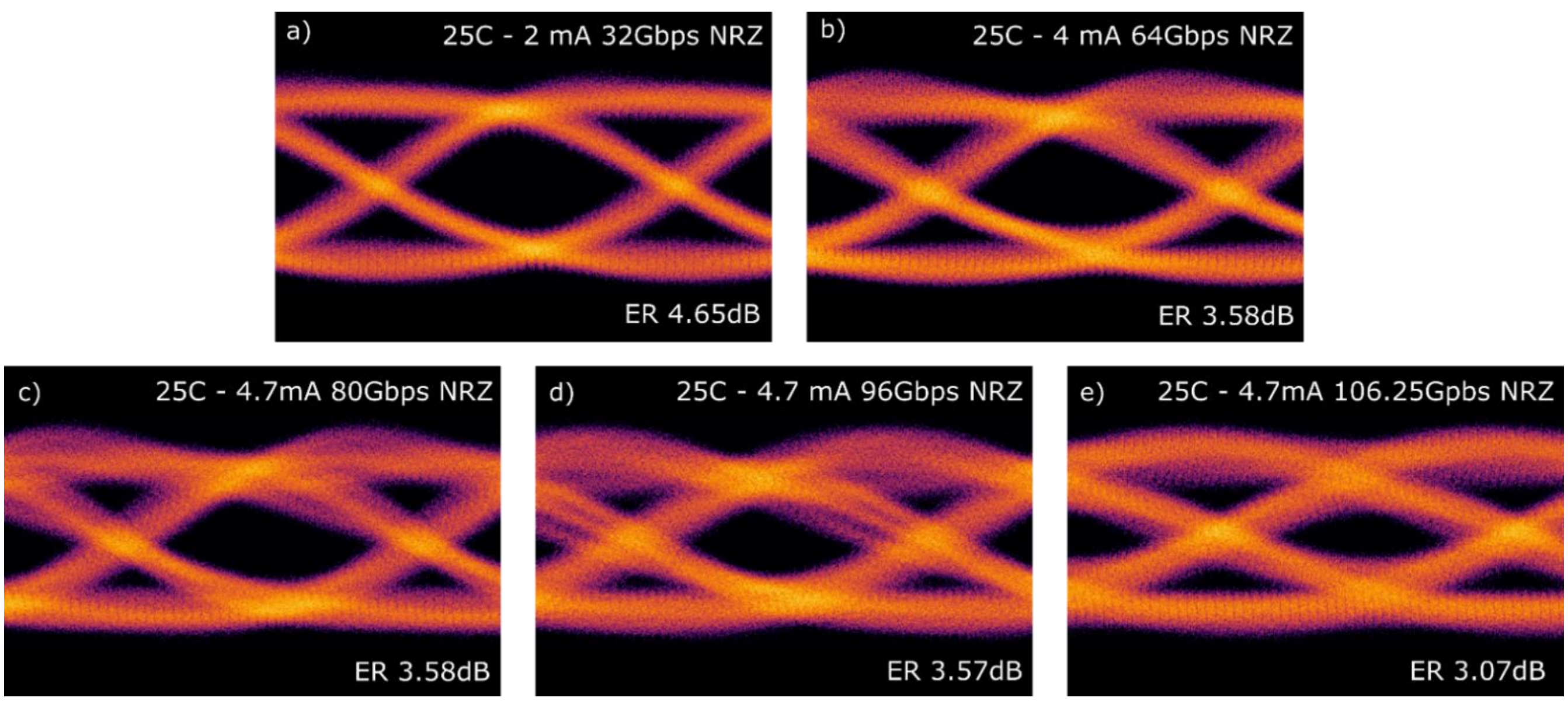

*Figure 5 Eye diagrams of the detuned-loading VCSEL under NRZ modulation at 25 °C for increasing data rates and bias currents. Measurements are shown at a) 32 Gb/s and 2 mA, b) 64 Gb/s and 4 mA, and c) 80, d) 96 and e) 106.25 Gb/s at 4.7 mA. The corresponding extinction ratios are 4.65, 3.58, 3.50, 3.57 and 3.07 dB, respectively.*

### *Large-signal modulation*

The large-signal measurements in Figure 5 demonstrate that the enhanced intrinsic dynamics of the detuned-loading VCSEL translate into high-speed data transmission over a broad range of operating conditions. Figure 5 shows NRZ eye diagrams measured in a back-to-back configuration using a PRBS13 sequence at progressively increasing data rates and bias currents. Already at a low bias current of only 2 mA, the device supports 32 Gb/s operation with a clearly open eye and an extinction ratio of 4.65 dB. Increasing the bias to 4 mA enables 64 Gb/s transmission with an extinction ratio of 3.58 dB. At an operating current of 4.7 mA and a bias voltage of 2.65 V, the device supports 80, 96, and ultimately 106.25 Gb/s, with extinction ratios of 3.50, 3.57, and 3.07 dB, respectively. The corresponding bit error rates (BERs), estimated from the eye diagrams (see Methods), are $8.5 \times 10^{-8}$, $1.8 \times 10^{-6}$, and $1.9 \times 10^{-4}$. These estimates remain below the KP-FEC threshold of $2.4 \times 10^{-4}$ up to 106.25 Gb/s. Importantly, 4.7 mA remains below the transverse-mode transition, so that the VCSEL operates in the engineered-mode regime where the detuned-loading mechanism is active. All eye diagrams shown in Figure 5 are obtained without digital signal processing, transmitter pre-emphasis, or receiver equalization, removing additional electronics that typically consume an order of magnitude more energy than the VCSEL itself [10][12]. The measured performance therefore directly reflects the intrinsic modulation capability of the device rather than compensation by the surrounding electronics. The ability of the same VCSEL to operate at moderate data rates at very low bias and to scale up to 106.25 Gb/s by increasing the operating current, while maintaining open eyes without electronic signal conditioning, highlights the broad usable operating range and the potential of transverse detuned loading for low-complexity, high-speed and energy efficient optical links.

## Conclusion

In conclusion, we have experimentally demonstrated transverse detuned loading in a directly modulated 850 nm single-mode VCSEL, showing that carrier-dependent transverse modal-loss engineering can substantially enhance the intrinsic laser dynamics over a broad temperature range. This approach enables a directly resolved carrier–photon relaxation frequency above 44 GHz and an electrical modulation bandwidth exceeding 50 GHz, supporting 106.25 Gb/s NRZ transmission without DSP, transmitter pre-emphasis, or receiver equalization.

Because transverse detuned loading relies on converting carrier-induced refractive-index changes into a favourable modulation of modal loss, the concept is not restricted to 850 nm or to a specific material platform. Its extension to other wavelengths and material systems, potentially combined with higher differential gain, offers a route toward still higher carrier–photon resonance frequencies and modulation bandwidths. More broadly, transverse modal-loss engineering provides a general degree of freedom for tailoring the dynamics of vertical-cavity lasers beyond conventional cavity designs.

## METHODS

### *A. Device design and fabrication*

The detuned-loading VCSEL is based on a conventional oxide-confined architecture with a modified upper cavity. The top semiconductor mirror contains six p-DBR pairs, and the surface of the semiconductor section is etched to form an intracavity surface relief close to the active region. Because this feature is located very close to the active region, it modifies the effective refractive index of the cavity and consequently alters the transverse refractive-index profile responsible for optical confinement. This provides an additional degree of freedom in the design of the modal properties of the VCSEL. The uppermost semiconductor layer consists of highly doped GaAs. This thin layer is selectively etched to realize the intracavity mode filter and serves two important functions. First, its high doping level enables the formation of a low-resistance metal contact, improving current injection and reducing parasitic effects during high-speed modulation. Second, due to its significant overlap with the standing-wave pattern inside the cavity, the unetched region introduces substantial optical absorption losses. As a result, a single design element simultaneously modifies the optical confinement and the spatial distribution of cavity losses. The cavity is completed by four dielectric DBR (dDBR) pairs composed of $Nb_2O_5$ and $SiO_2$, providing the required mirror reflectivity. The final layer is a quarter-wave-thick silicon layer. This layer is also selectively etched to restore the original mirror reflectivity in the central region of the device. Where the silicon remains, the phase condition at the mirror surface significantly reduces the local reflectivity, thereby increasing the mirror losses. Consequently, two distinct loss mechanisms are implemented: the intracavity relief modifies both the refractive-index and absorption-loss profiles, while the silicon layer shapes the mirror-loss distribution.

### *B. Numerical modelling*

Static COMSOL simulations were used to evaluate how refractive-index changes in the pumped MQW region modify the transverse modes. The 2.5D model includes the refractive-index step introduced by the intracavity filter together with the spatially non-uniform loss distribution. The COMSOL model does not explicitly couple the thermal, electrical, and optical physics. Their influence is instead approximated through effective modifications of the MQW refractive-index distribution, providing a first-order representation of these effects while keeping the model computationally tractable. The MQW refractive index of the pumped region is swept to represent two effects. A larger positive variation describes the quasi-static evolution associated with thermal lensing as the bias current increases; in the simplified model, the range is estimated from an experimentally observed wavelength shift of approximately 3 nm and an optical confinement factor of 1%. A much smaller negative perturbation represents the carrier-induced refractive-index change during high-speed modulation. The large-scale sweep therefore describes the bias-dependent steady-state mode profile, while the local derivative of modal loss with respect to MQW refractive index around each operating point provides the relevant dynamic modal-loss sensitivity. The eigenmodes and their losses are calculated and then compared with measured near-fields and spectral evolution.

The simulated loss slope is then used to estimate the imaginary part of the reactive confinement factor, which represents how strongly carrier-induced refractive-index modulation changes the modal loss. From Eq. 4 we can calculate the maximum enhancement of the relaxation frequency as

$$\sqrt{Re\left[(1+i\alpha)\tilde{\Gamma}\right]} \leq \sqrt[4]{1+\alpha^2}\sqrt{\left|\tilde{\Gamma}\right|}$$

Assuming $\alpha$=4 [1] and $\left|\tilde{\Gamma}\right|$ to be approximately equal for the detuned and undetuned VCSELs, we obtain a maximum possible enhancement of around 100%.

### *C. Measurements and analysis*

For the small signal measurements, the fabricated device was mounted on a temperature-controlled sample holder. The emitted light was collected using a lensed fibre. The electrical modulation signal was combined with the DC bias using a bias tee and applied to the device through an RF ground–signal–ground (GSG) probe.

The small-signal electro-optical response of the 850 nm VCSEL was measured in a 50 Ω configuration up to 50 GHz using a Keysight N5225B PNA vector network analyser together with a Keysight N4373E Lightwave component analyser. The transmission losses of the electrical and optical components including those of the probe, connectors and photodetector were compensated by subtracting their $S21$ responses. To estimate $f_{res}$, $\gamma_d$ and 3dB bandwidths exceeding our 50 GHz measurement range, we fit the de-embedded response to a three-pole transfer function, $H(f)$, in which a single real pole was found sufficient to accurately describe the measured parasitic roll-off.

$$H(f) = \frac{f_r^2}{(f_r^2 - f^2 + i\gamma_d f/2\pi)\left(1 + i\frac{f}{f_p}\right)}$$

Here, $f, f_r$ and $f_p$ are the modulation, relaxation, and parasitic frequencies, respectively and $\gamma_d$ is the relaxation damping. Measurements were performed at heatsink temperatures of 25 °C, 50 °C and 75 °C using the same analysis.

For the large-signal measurements, the device was mounted on a sample holder at room temperature, and the emitted light was collected using a lensed fiber. An arbitrary waveform generator (AWG) generated the electrical data signal, which was combined with the DC bias using a bias tee and applied to the device through an RF ground–signal–ground (GSG) probe. The optical output was recorded using a digital communications analyzer (DCA) to obtain eye diagrams and determine the extinction ratios. Bit error rates (BERs) were estimated from the eye diagrams. The pre-FEC BER was estimated from the statistical distributions of the logical-zero and logical-one levels in the measured eye diagram. The quality factor was calculated as $Q = (\mu_1 - \mu_0)/(\sigma_1 + \sigma_0)$, where μ and σ denote the mean and standard deviation of each level, respectively. Assuming Gaussian noise, the BER was estimated as $\mathrm{BER} = \frac{1}{2}\mathrm{erfc}\left(Q/\sqrt{2}\right)$ [2].

## ACKNOWLEDGMENTS

This work was partially supported by the Engineering and Physical Research Council and PIADS (Grant number EP/S023321/1).

## DATA AVAILABILITY

The data that support the findings of this study are available from the corresponding author upon reasonable request.